\documentclass[aps,pra,preprint,showpacs,amsmath,amsfonts,amssymb,floatfix]{revtex4}
\usepackage{amsmath,amsfonts,amssymb,color}
\usepackage{leftidx}
\usepackage{graphicx}
\usepackage{dcolumn}
\usepackage{bm}
\usepackage{epstopdf}
\usepackage{epsfig}

\begin{document}
\title{Effects of Dephasing on Quantum Adiabatic Pumping with Nonequilibrium Initial States}
\author{Longwen Zhou, Da Yang Tan, and Jiangbin Gong}
\affiliation{Department of Physics, National University of Singapore, Singapore 117546}
\date{\today}


\begin{abstract}
Thouless's quantum adiabatic pumping is of fundamental interest to condensed-matter physics. It originally considered a zero-temperature equilibrium state
uniformly occupying all the bands below a Fermi surface.  In the light of
recent direct simulations of Thouless's concept in cold-atom systems,
this work investigates the dynamics of quantum adiabatic pumping subject to
dephasing, for rather general initial states with nonuniform populations and possibly interband coherence.
Using a theory based on pure-dephasing Lindblad evolution, we find that
the pumping is contributed by two parts of different nature, a dephasing-modified geometric part weighted by initial Bloch state populations, and an interband-coherence-induced part
compromised by dephasing, both of them being independent of the pumping time scale.
The overall pumping reflects an interplay of the band topology, initial state populations, initial state  coherence, and dephasing. Theoretical results are carefully checked in a Chern insulator model coupled to a pure-dephasing environment, providing a useful starting point to understand and coherently control quantum adiabatic pumping in general situations.

\end{abstract}


\maketitle

{\it Introduction.}
Thouless's seminal work on adiabatic quantum pumping~\cite{Thouless1983,Niu} establishes a deep connection between band topology and quantum transport. In particular, the sum of the Chern numbers of all the filled bands below the Fermi surface determines the number of pumped charges over one adiabatic cycle in a one-dimensional~($1$D) periodic lattice. To date various quantum pumps have been proposed to study topological phases and  topological phase transitions~\cite{Berg,ChenSpinPump,Chong,Cooper,Dai,Kraus,Kwek,Marra,Mueller,Vanderbilt}. Thouless's concept, which can be deemed as a dynamical version of the integer quantum Hall effect, is directly simulated this year in two cold-atom experiments~\cite{exp1,exp2}. Thouless's pumping has also been extended to periodically driven quantum systems to manifest the topology of Floquet quasi-energy bands instead of energy bands~\cite{DerekPRL,Longwen,Hailong}.

In simulating quantum pumping with cold-atom~\cite{exp1,exp2} and waveguide~\cite{Kraus} systems, the initial state might not be the zero-temperature equilibrium state considered by Thouless. There are at least three reasons why nonequilibrium initial states should be investigated theoretically. First, if a bosonic system is pumped, there does not exist a Fermi surface to automatically guarantee full and uniform band filling. Second, even for a fermionic system, loading the particles into a lattice in an actual experiment is nontrivial and might not result in uniform populations within one band~\cite{exp2}. Third, a simple initial state localized in an optical or waveguide lattice may possess interband coherence~(IBC)~\cite{Hailong} in the band representation. These nonequilibrium situations occur even more frequently in studies of quantum adiabatic pumping in periodically driven systems~\cite{DerekPRL,Hailong,Longwen}. It is hence of both fundamental and practical interest to extend Thouless's pumping to nonequilibrium initial states. Indeed, in a recent study, IBC is found to induce a remarkable correction to adiabatic pumping in periodically driven systems~\cite{Hailong}.

In this Letter we study dephasing effects on adiabatic pumping dynamics with rather general initial states. On the one hand, realistic systems are always subject to dephasing, so how dephasing affects quantum pumping is of theoretical interest. On the other hand, in well controlled experimental systems, it is possible to deliberately introduce dephasing effects into the setup so as to manifest the findings described below. Specifically, with a minimal pure-dephasing model and certain assumptions, we find that the number of pumped particles over one cycle comprises two components of different nature: one dephasing-modified geometric component weighted by initial populations on different Bloch states, and a second component induced by IBC but compromised by dephasing, with both of them independent of the pumping time scale.  
The overall pumping hence depends on the band topology, initial Bloch state populations, initial IBC and dephasing, offering a stimulating starting point to understand quantum adiabatic pumping with dephasing and in nonequilibrium situations. Our theory is fully checked in a Chern insulator model coupled to a pure-dephasing environment.

{\it Dephasing-induced correction to nonadiabatic population transfer.}
Our theory starts from a slowly varying Hamiltonian $H(s)$, where $s=vt$~(it is straightforward to extend to other time dependence of $s$) represents a tunable system parameter, $t$ is the time variable, and $v$ is the sweeping rate of $s$. All variables are assumed to be scaled and dimensionless. The Planck constant $\hbar$ is set to unity throughout. For convenience $s=vt$ is also used to reflect the time $t=s/v$ when appropriate. An adiabatic protocol starts at $s=s_0$ and ends at $s=s_1$. We assume that for any value of $s\in[s_0,s_1]$, the spectrum of $H(s)$ is never degenerate.  Hence the conventional quantum adiabatic theorem applies when the sweeping rate $v\rightarrow0$. For a small but finite value of $v$, the lowest-order nonadiabatic corrections should be carefully studied in order to capture the main physics of adiabatic pumping.

To account for dephasing in a solvable manner in the context of adiabatic pumping, we exploit the following master equation in the Lindblad form~\cite{Breuerbook}: $v\frac{\rm{d}}{{\rm{d}}s}\rho(s)=-i[H(s)\rho(s)-\rho(s)H(s)]
+\gamma\left\{A(s)\rho(s)A(s)-\frac{1}{2}\left[A^2(s)\rho(s)+\rho(s)A^2(s)\right]\right\}$. Here $s$ is already used to reflect the time, $\rho(s)$ is the density matrix of the system at $t=s/v$, $A(s)$ is a Hermitian Lindblad operator, and $\gamma$ denotes the dephasing rate. Though this treatment does not explicitly consider bath degrees of freedom and has ignored all non-Markovian effects, it does allow us to closely follow our previous work~\cite{Hailong} and to clearly reveal the competition between IBC and dephasing. Further, to have a simplest model incorporating dephasing effects,
the time-varying Hamiltonian $H(s)$ and the Lindblad operator $A(s)$ are assumed to be always commutable. By this construction, the system-environment coupling does not directly introduce transitions between instantaneous eigenstates of $H(s)$. That is, we adopt a model based on pure-dephasing Lindblad evolution~\cite{AvronLZT}. The instantaneous eigenstates of $H(s)$ are denoted by $|m(s)\rangle$, with $H(s)|m(s)\rangle=E_m(s)|m(s)\rangle$.
Because $A(s)$ commutes with $H(s)$, $|m(s)\rangle$ are also chosen as eigenstates of $A(s)$, with $A(s)|m(s)\rangle=A_m(s)|m(s)\rangle$.

We first aim to show how both nonadiabaticity and dephasing jointly change the populations on, and the coherence between, instantaneous eigenstates of $H(s)$, up to the first order in $v$ during a pumping protocol. We project $\rho(s)$ onto the instantaneous eigenstates $\langle m(s)|$ and $|j(s)\rangle$ of $H(s)$, yielding $\rho(s)=\sum_{mj}\rho_{mj}(s)|m(s)\rangle \langle j(s)|$. This time-dependent representation for system's density matrix will be used throughout. To the first order in $v$, we obtain the off-diagonal elements~(coherence)~(see Appendix~\cite{notesupp})
\begin{equation}
 \rho_{mj}(s)=v\frac{\dot{\rho}_{mj}(s)+[\rho_{jj}(s)-\rho_{mm}(s)]\langle m(s)|\frac{{\rm d}}{{\rm d}s}|j(s)\rangle}{-i[E_{m}(s)-E_{j}(s)]-\frac{\gamma}{2}[A_{m}(s)-A_{j}(s)]^{2}},
\label{PDLPapp0Eq4-text}
\end{equation}
as well as the diagonal elements~(populations)
\begin{eqnarray}
&&\rho_{jj}(s_1)-\rho_{jj}(s_0) \nonumber \\
&=& -v\sum_{m\ne j}\left[\rho_{mj}(s)\frac{\langle j(s)|\frac{{\rm{d}}H(s)}{{\rm{d}}s}|m(s)\rangle}{g_{mj}(s)[\gamma_{mj}(s)+ig_{mj}(s)]}\bigg|_{s=s_0}+{\rm c.c.} \right] \nonumber
        \\
&                     & - 2v\sum_{m\ne j}[\rho_{jj}(s_0)-\rho_{mm}(s_0)]\int_{s_0}^{s_1} B(s)\ {\rm{d}}s
\label{finalTsp}
\end{eqnarray}
where $B(s)\equiv\big|\langle j(s)|\frac{{\rm{d}}H(s)}{{\rm{d}}s}|m(s)\rangle\big|^{2}\frac{\gamma_{mj}(s)}{g^2_{mj}(s)\left[\gamma^2_{mj}(s)+g^2_{mj}(s)\right]}$,
$g_{mj}(s)\equiv E_m(s)-E_j(s)$, $\gamma_{mj}(s)\equiv \frac{\gamma}{2}\left[A_m(s)-A_j(s)\right]^2$. It is enlightening to discuss the nonadiabatic transition probabilities in Eq.~(\ref{finalTsp}). There it is seen that $\Delta P_j\equiv \rho_{jj}(s_1)-\rho_{jj}(s_0)$  is composed of two parts: one part related to initial state coherence $\rho_{mj}(s_0)$, and a second part depending on the differences in the initial populations on eigenstates of $H(s_0)$. The coherence properties of the final density matrix do not appear because $\rho_{mj}(s_1) (m\ne j)$,  the off-diagonal terms of $\rho(s_1)$, have been washed off by dephasing. Interestingly, the initial-state-coherence induced correction to $\Delta P_j$ does persist in the presence of dephasing, with the gap function $g_{mj}(0)$ however replaced by $g_{mj}(0)-i\gamma_{mj}(0)$. Furthermore, the second part of $\Delta P_j$ shows that, even without initial state coherence, dephasing alone can also induce a nonadiabatic correction proportional to $v$. As such, both initial state coherence and dephasing can affect the proportionality coefficient $\Delta P_j/v$, a clear interplay expected to influence strongly on adiabatic pumping. Note also that if only one single eigenstate of $H(s_0)$ is populated at $s=s_0$, then the above result is fully consistent with the one derived by Avron \emph{et al} for a pure-dephasing Landau-Zener model~\cite{AvronLZT}.  Numerical results in two- and three-level systems fully confirm the theoretical results here~\cite{notesupp}.

{\it Dephasing-modified adiabatic pumping.}
In adiabatic pumping, nonadiabatic corrections of the order $v$ could accumulate over a pumping cycle, potentially yielding an overall outcome independent of $v$ or the pumping time scale~\cite{Hailong}. Consider then a particle moving in a $1$D periodic lattice, subject to pure-dephasing Lindblad evolution. The Hamiltonian is assumed to be $H_k(s)$, with $k\in [-\pi,\pi)$ being the quasimomentum and $H_k(s)=H_k(s+2\pi)$. The parameter space formed by $k$ and $s$ is hence a $2$-dimensional~($2$D)~torus. The eigenvalues of $H_k(s)$ are assumed to form well-gapped Bloch bands. A pumping cycle can be realized by sweeping $s$ slowly from $s_0=0$ to $s_1=2\pi$ with a speed $v$. The Lindblad dephasing operators for the $k$-component of the system are assumed to be $A_k(s)$, which preserves the translational invariance. Due to this treatment, $k$ is conserved during the pumping and states with different $k$ are always independent of each other. Therefore, Eqs.~(\ref{PDLPapp0Eq4-text}) and~(\ref{finalTsp}) are applicable for each individual $k$ value. As elaborated in Appendix, the spectrum of $H_k(s)$ and $A_k(s)$, as well as the initial state populations of the system, are all assumed to have reflection symmetries in the $k$-space so as to highlight all the pumping terms that are independent of $v$.

The number of particles pumped through per unit cross section at the end of the pumping cycle can be written as
a $2$D integral, namely, $Q=\frac{1}{2\pi}\int_{-\pi}^{\pi}{\rm{d}}k\int_{0}^{2\pi}{\rm {d}}s\ f_{ks}$, with
\begin{equation}
f_{ks} \equiv {\rm{Tr}}\left[v^{-1}\rho(s)v_k(s)\right],
\label{PCurrent}
\end{equation}
where $v_k(s)\equiv\frac{{\rm d}H_k(s)}{{\rm d}k}$ is the group velocity operator of a particle with quasimomentum $k$.  To evaluate $f_{ks}$, both the diagonal and off-diagonal elements $\rho_{jj}(s)$ and $\rho_{mj}(s)$ are needed. Indeed,
$f_{ks}=v^{-1}\sum_j\rho_{jj}(s)\langle j(s)|v_{k}(s)|j(s)\rangle + v^{-1}\sum_{m\ne j}\rho_{mj}(s)\langle j(s)|v_k(s)|m(s)\rangle$. A straightforward but somewhat tedious application of Eqs.~(\ref{PDLPapp0Eq4-text}) and~(\ref{finalTsp})~\cite{notesupp} then yields~(to the first order of $v$) four subterms of $f_{ks}$, i.e., $f_{ks}=f_{ks}^{(a)} + f_{ks}^{(b)}+ f_{ks}^{(c)}+ f_{ks}^{(d)}$, with
\begin{eqnarray}
f_{ks}^{(a)}&=&2{\sum_{\substack{j,m\\m\ne j}}}\rho_{jj}(0)
\frac{\Gamma_{mj}(s){\rm Re}\left[\langle \frac{{\rm d}j(s)}{{\rm d}k}|m(s)\rangle\langle m(s)|\frac{{\rm d} j(s)}{{\rm d}s}\rangle\right]}{\Gamma^2_{mj}(s)+1} \nonumber \\
f_{ks}^{(b)}&=&2{\sum_{\substack{j,m\\m\ne j}}}\rho_{jj}(0) \frac{{\rm Im}\left[\langle \frac{{\rm d} j(s)}{{\rm d}k} |m(s)\rangle\langle m(s)|\frac{{\rm d}j(s)}{{\rm d}s}\rangle\right]}{\Gamma^2_{mj}(s)+1}
\nonumber \\
f_{ks}^{(c)}&=&-{\sum_{\substack{j,m\\m\ne j}}} 2\frac{\partial E_j(s)}{\partial k}{\rm Re}\left[\rho_{mj}(0)C(s=0)\right]
 \nonumber \\
f_{ks}^{(d)}&=&\frac{1}{2\pi}{\sum_{\substack{j,m\\m\ne j}}}\rho_{mj}(0)\frac{\langle j(0)|\frac{{\rm d}H(0)}{{\rm{d}}k}|m(0)\rangle}{\gamma_{mj}(0)+ig_{mj}(0)}, \nonumber \\
\label{TCfinal}
\end{eqnarray}
where $C(s)\equiv\frac{\langle j(s)|\frac{{\rm d}H(s)}{{\rm{d}}s}|m(s)\rangle}{g_{mj}(s)[\gamma_{mj}(s)+ig_{mj}(s)]}$, $\Gamma_{mj}(s)\equiv\frac{\gamma_{mj}(s)}{g_{mj}(s)}$, and all other quantities do carry a $k$-dependence but not spelled out explicitly, For example, $\rho_{mj}(s)$ now means the density matrix element already projected onto the subspace with quasimomentum $k$, $|j(s)\rangle$ above refers to an eigenstate of $H_k(s)$, $\Gamma_{mj}(s)$ should also be understood as a function of $k$, with $\gamma_{mj}(s)$ and $g_{mj}(s)$ defined at the same individual $k$ values. As seen above, all the four subterms of $f_{ks}$ are now independent of $v$. That is, accumulating the transport behavior over one entire pumping cycle allows us to capture all the subtle nonadiabatic and dephasing effects proportional to $v$. Equation~(\ref{TCfinal}) goes beyond a previous result~\cite{AvronDQR} because we not only incorporate interband dephasing (as reflected by $\Gamma_{mj}$) as a function of the band gaps, but also account for IBC.

Each of the four subterms in Eq.~(\ref{TCfinal}) should be discussed. We do so mainly by asking what happens to them
if removing dephasing~[$\gamma_{mj}(s)=0$, $\Gamma_{mj}(s)=0$]. First, $f_{ks}^{(a)}$ depending on initial populations would vanish, hence a subterm entirely due to dephasing. Second, $f_{ks}^{(b)}$ expectedly reduces to an integral over Berry curvatures of Bloch wavefunctions weighted by their initial populations. Third, $f_{ks}^{(c)}$ would recover an expression parallel to that in Ref.~\cite{Hailong}, where how IBC in driven systems corrects adiabatic pumping was first studied. This is especially encouraging because our derivation here refers to a nondriven system and also takes an entirely different route than Ref.~\cite{Hailong}. The alternative derivation in Ref.~\cite{Hailong} clearly indicates that $f_{ks}^{(c)}$ is of a dynamical, not a geometrical origin. The found IBC effect is seen to persist well in the presence of dephasing, insofar as its explicit form is only modified by dephasing via $g_{mj}(s)\rightarrow g_{mj}(s)-i\gamma_{mj}(s)$~[see the expression of $C(s)$ defined below Eq.~(\ref{TCfinal})]. Lastly, in obtaining $f_{ks}^{(d)}$ we already assumed that the off-diagonal elements $\rho_{mj}(s_1=2\pi)$ have decayed to zero~\cite{notesupp}. Hence naively setting $\gamma_{mj}(0)=0$ alone in the expression for $f_{ks}^{(d)}$ does not suffice to recover any unitary limit. Because this last subterm does not even depend on the pumping protocol~(no derivatives with respect to $s$ is involved), this subterm largely originates from a current inherent in the initial state itself, whose contribution to the pumping is accumulated only for a dephasing time scale. Overall, the four subterms presented in Eq.~(\ref{TCfinal}) can now be used to predict, both qualitatively and quantitatively, the features of adiabatic transport emanating from rather general nonequilibrium states.


{\it Adiabatic pumping in a Chern-insulator model.}
We use the Qi-Wu-Zhang model Hamiltonian~\cite{zhang},
\begin{equation}
H_k(s)=\sin(k)\sigma_x+\sin(s)\sigma_y+[\delta +\cos(k)+\cos(s)]\sigma_z,
\label{QWZM}
\end{equation}
which contains the main feature of Chern insulators~\cite{BernevigBook}. This system can describe spin-$1/2$~($\sigma_{x,y,z}$ are Pauli matrices) fermions with spin-dependent nearest-neighbor hoppings on a square lattice, with $\delta$ being an energy bias parameter. Though originally $k$ and $s$ refer to two quasimomenta along different directions, here we may also view the above Hamiltonian as a $1$D system with $s$ being an experimentally tunable system parameter. The instantanenous eigenstates of $H_k(s)$ are represented as $|\psi_{k}^{(1)}(s)\rangle$ and $|\psi_{k}^{(2)}(s)\rangle$, with eigenvalues $E_{k,1}(s)<E_{k,2}(s)$. To introduce pure dephasing, we assume the Lindbald operators $A_k(s)$ to be the same as $H_k(s)$. The spectrum of $H_k(s)$~[hence also of  $A_k(s)$] is indeed symmetric in $k$. Remarkably, Eq.~(\ref{TCfinal}) applied to this two-band model yields particularly simple expressions for $f_{ks}^{(a)}$ and $f_{ks}^{(b)}$, with
\begin{eqnarray}
f_{ks}^{(a)}&=&\Delta \rho_{k}(0)\frac{\Gamma_{21}}{\Gamma_{21}^{2}+1}G_{ks},\nonumber \\
f_{ks}^{(b)}&=&\Delta \rho_{k}(0)\frac{1}{\Gamma_{21}^{2}+1}\Omega_{ks},
\label{fab}
\end{eqnarray}
where $\Gamma_{21}$, as defined before, is now given by $\Gamma_{21}=\gamma\left[E_{k,2}(s)-E_{k,1}(s)\right]/2$. $\Delta\rho_{k}(0)$ in Eq.~(\ref{fab}) refers to the initial population difference between the ground and excited bands. Here $ G_{ks}\equiv 2{\rm Re}\left[\langle\frac{{\rm d}}{{\rm d}k}\psi^{(1)}_k(s)|\psi^{(2)}_k(s)\rangle\langle\psi^{(2)}_k(s)|\frac{{\rm d}}{{\rm d}s}\psi_k^{(1)}(s)\rangle\right]$ and $\Omega_{ks}\equiv 2{\rm Im}\left[\langle\frac{{\rm d}}{{\rm d}k}\psi^{(1)}_k(s)|\frac{{\rm d}}{{\rm d}s}\psi_{k}^{(1)}(s)\rangle\right]$ are the Fubini-Study metric of the ground state bundle and the more familiar Bloch band Berry curvature~\cite{AvronDQR}. Clearly then, without IBC, the adiabatic pumping solely determined by $f_{ks}^{(a)}$ and $f_{ks}^{(b)}$ is of geometrical nature, but modified by dephasing. The whole story of adiabatic transport under a pumping protocol is however completed by $f_{ks}^{(c)}$ and $f_{ks}^{(d)}$, whose explicit expressions follow Eq.~(\ref{TCfinal}) and are not given here.

\begin{figure} 
\begin{center}
\resizebox{0.65\textwidth}{!}{%
  \includegraphics[trim=0.6cm 1.0cm 0.5cm 9.0cm, clip=true, height=!,width=15cm]{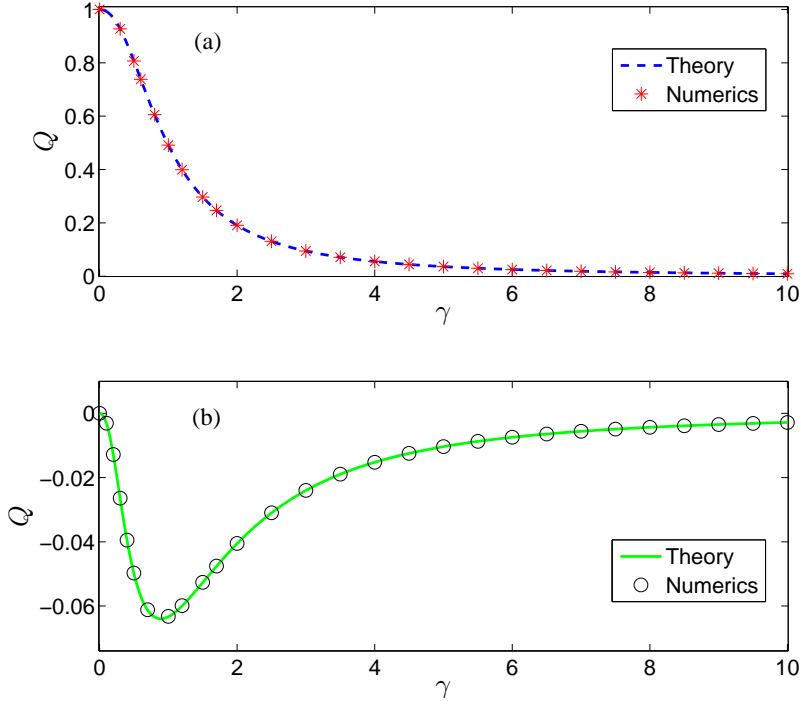}
}
\end{center}
\caption{(color online) Number of pumped particles $Q$ vs. the dephasing rate $\gamma$ for~(a).~$\delta=1.0$, (b).~$\delta=2.5$. The system Hamiltonian is described in Eq.~(\ref{QWZM}). The adiabatic sweeping rate is assumed to be $v=10^{-3}$ in our numerical calculations.}
\label{fig:PCIC}
\end{figure}

To verify our theory, we numerically evolve various initial states at each quasimomentum value using the above-mentioned pure-dephasing Lindblad master equation. We then obtain $Q$ by integrating the numerically found $f_{ks}$ over $k$ and $s$. Consider first an initial state uniformly filling the bottom band only. In this case there is no IBC, so the pumping is entirely determined by $f_{ks}^{(a)}$ and $f_{ks}^{(b)}$ in Eq.~(\ref{fab}). As shown in Fig.~\ref{fig:PCIC}, theory and numerics are in perfect agreement. For the topologically nontrivial case with $\delta=1$~(top panel), $Q$ decreases from a quantized Chern number~($Q=1$) to zero as the dephasing rate $\gamma$ increases. Thus, the topological relevance to adiabatic pumping is gradually suppressed by dephasing. For the topological trivial case with $\delta=2.5$~(bottom panel), however, $Q$ first decreases and then increases with $\gamma$. We stress that this dephasing-induced pumping is still directly connected with the $G_{ks}$ metric.
 The non-monotonous dependence of $Q$ on the dephasing rate $\gamma$
 is easily understandable from the expression of $f_{ks}^{(a)}$, which
 reflects a competition between the dephasing-modified (effective) band gaps and dephasing-modified nonadiabatic transition rates.

\begin{figure} 
\begin{center}
\resizebox{0.65\textwidth}{!}{%
  \includegraphics[trim=0.6cm 1.0cm 0.5cm 9.0cm, clip=true, height=!,width=15cm]{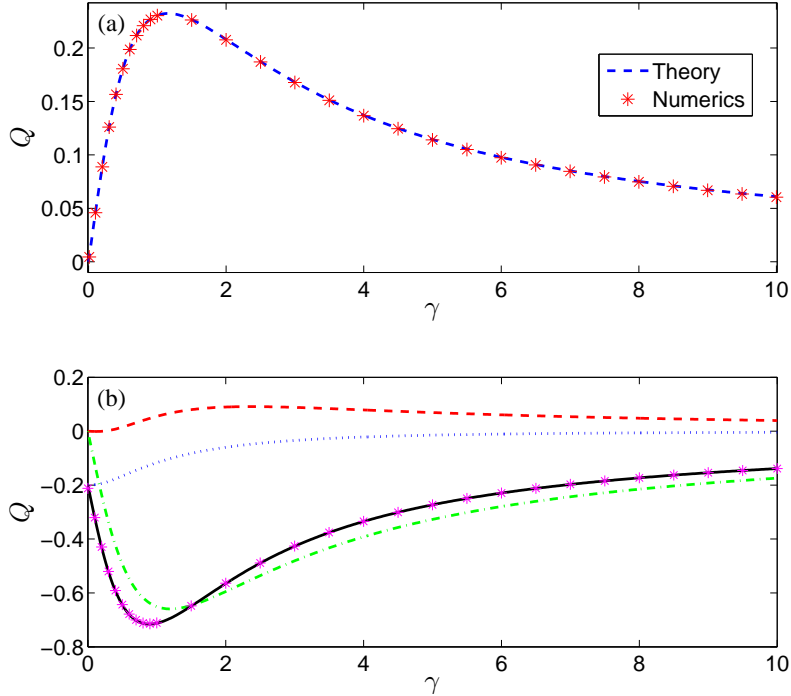}
}
\end{center}
\caption{(color online) Number of pumped particles $Q$ vs $\gamma$. In (a), the initial state equally populates the two bands with IBC and in (b), the initial state unequally populates the two bands with IBC. See the main text for details of the initial states used. The system Hamiltonian is described in Eq.~(\ref{QWZM}), with $\delta=-0.5$ in (a) $\delta=-1.6$ in (b). The adiabatic sweeping speed is chosen to be $v=10^{-3}$ in our numerics. In (b),~(blue) dotted line, (green) dashed line, and (red) dot-dashed line represent the respective contribution from $f_{ks}^{(b)}$, $f_{ks}^{(c)}$, and $f_{ks}^{(d)}$ in Eq.~(\ref{TCfinal});  the black line represents the total pumping as compared with the numerical results represented by discrete points.}
\label{fig:PCC}
\end{figure}

We next consider initial states that coherently populate the two bands. As the first example, the initial density matrix is chosen as $\frac{1}{2}\left[|\psi^{(1)}_k(0)\rangle + |\psi_k^{(2)}(0)\rangle\right] \otimes\left[\langle\psi_k^{(1)}(0)|+\langle\psi_{k}^{(2)}(0)|\right]$. This initial state populates the two bands equally with a uniform distribution in $k$, with $\Delta\rho_{k}(0)=0$. So $f_{ks}^{(a)}$ and $f_{ks}^{(b)}$ described in Eq.~(\ref{fab}) have no contributions to adiabatic pumping. Nevertheless, the effect of IBC on the transport is nonzero, with the agreement between theory and numerics presented in Fig.~\ref{fig:PCC}(a). It is seen that as dephasing strengthens, $Q$ increases first and then decreases, reflecting a competition between dephasing and IBC. In particular, the IBC induced transport is significant even when the dephasing rate $\gamma$ is comparable to the characteristic scale of the system's band gap. Turning to a second density matrix as the initial condition, i.e., $\left[\sqrt{0.6}|\psi^{(1)}_k(0)\rangle+\sqrt{0.4}e^{ik}|\psi_k^{(2)}(0)\rangle_k\right]$ $\otimes$ $\left[\sqrt{0.6}\langle\psi_k^{(1)}(0)|+\sqrt{0.4}e^{-ik}\langle\psi_k^{(2)}(0)|\right]$. The $f_{ks}^{(a)}$ term for this case turns out to make no contribution. Both theoretical and numerical results for this example are shown in Fig.~\ref{fig:PCC}(b). There, we have separately plotted the contributions associated with $f_{ks}^{(b)}$, $f_{ks}^{(c)}$, and $f_{ks}^{(d)}$. Though the overall dependence of $Q$ on $\gamma$ shows one valley only, it is seen that each of the three terms responds to dephasing in different manners. That is, it is necessary to know all these terms in order to better understand the overall pumping.

{\it Concluding remarks.}
Quantum adiabatic pumping with rather general nonequilibirum initial states and in the presence of dephasing is contributed by two components of different nature, with one of them depending on initial Bloch state populations and the other determined by interband coherence. Though our theory uses a very simple pure-dephasing Lindblad master equation, the explicit results obtained here should be useful towards understanding the dynamics of quantum pumping in more realistic situations. By using different initial states or different pumping protocols, it is possible to isolate each of the two pumping components. We highlight the interband coherence effect, as it persists well in the presence of dephasing. Note that the interband coherence effect on pumping~[$f_{ks}^{(c)}$ in Eq.~(\ref{TCfinal})] is determined by $\frac{{\rm\ d}H(s)}{{\rm d}s}$ at the start, so its magnitude can be extensively manipulated by varying the switch-on behavior of a pumping protocol. This promises an interesting and relatively robust means to coherently control the pumping dynamics albeit dephasing.

\vspace{0.5cm}
{\bf Acknowledgments:} We thank Hailong Wang for helpful discussions.


\appendix

\vspace{1cm} \begin{center} {\bf Appendix} \end{center}

This Appendix has three sections. In Appendix A, we present detailed derivations of Eq.~(1) and Eq.~(2) in the main text. This is followed by Appendix B, where numerical results are described and compared with theoretical results. The most important section is perhaps Appendix C, which gives a detailed derivation of the four subterms of $f_{ks}$ used in the main text. Whenever possible, we use the same notation as in the main text.

\section{ Details about $\rho_{mj}(s)$ and $\rho_{jj}(s)$}
\label{PDLPapp0}

We first present again the Lindblad master equation used in the main text,
\begin{eqnarray}
v\frac{\rm{d}}{{\rm{d}}s}\rho(s)&=&-i\,\left[H(s)\rho(s)-\rho(s)H(s)\right]\nonumber \\
&+&\gamma\,\left\{A(s)\rho(s)A(s)-\frac{1}{2}\left[A^2(s)\rho(s)+\rho(s)A^2(s)\right]\right\}.
\label{Leq}
\end{eqnarray}
Projecting both sides of this master equation onto $|m(s)\rangle$ and $\langle j(s)|$, the instantaneous eigenstates of $H(s)$, one obtains
\begin{eqnarray}
v\,\langle j(s)|\frac{{\rm d}}{{\rm d}s}\rho(s)|m(s)\rangle&=&-i\,\langle j(s)|H(s)\rho(s)-\rho(s)H(s)|m(s)\rangle \nonumber \\
&+&\ \gamma\, \langle j(s)|A(s)\rho(s)A(s)-\frac{1}{2}\left[A^{2}(s)\rho(s)+\rho(s)A^2(s)\right]|m(s)\rangle
\label{PDLPapp0Eq1}
\end{eqnarray}
Note that if $j=m$, then all the terms on the right hand side~(RHS) of Eq.~(\ref{PDLPapp0Eq1}) will disappear. As introduced in the main text, $\rho(s)$ is now expressed in the representation of instantaneous energy eigenstates of
$H(s)$, with $\rho(s)=\sum_{l,n}\rho_{ln}(s)|l(s)\rangle\langle n(s)|$. Plugging this expansion of $\rho(s)$ into Eq.~(\ref{PDLPapp0Eq1}), we find the diagonal density matrix elements
\begin{equation}
\dot{\rho}_{jj}(s)=-\sum_{m\neq j}\left[\rho_{mj}(s)\langle j(s)|\frac{{\rm d}}{{\rm d}s}|m(s)\rangle+{\rm c.c.}\right].
\label{PDLPapp0Eq2}
\end{equation}
Here ``$\cdot$" means the derivative with respect to the system parameter $s$. For an adiabatic process starting at $s=s_0$ and ending at $s=s_1$, the final population on level $j$ is then given by
\begin{equation}
\rho_{jj}(s_{1})=\rho_{jj}(s_{0})-\sum_{m\neq j}\int_{s_{0}}^{s_{1}}\left[\rho_{mj}(s)\langle j(s)|\frac{{\rm d}}{{\rm d}s}|m(s)\rangle+{\rm c.c.}\right].
\label{PDLPapp0Eq3}
\end{equation}
In a similar manner, letting $m\neq j$ in Eq.~(\ref{PDLPapp0Eq1}), we find the off-diagonal density matrix elements,
\begin{equation}
\rho_{mj}(s)=v\frac{\langle m(s)|\dot{\rho}(s)|j(s)\rangle}{-i\left[E_m(s)-E_j(s)\right]-\frac{\gamma}{2}\left[A_m(s)-A_j(s)\right]^2}.
\label{Ode}
\end{equation}
The numerator in Eq.~(\ref{Ode}) can be further rewritten as
\begin{equation}
\begin{split}\langle m(s)|\dot{\rho}(s)|j(s)\rangle=\dot{\rho}_{mj}(s)+\left[\rho_{jj}(s)-\rho_{mm}(s)\right]\langle m(s)|\frac{{\rm {\rm d}}}{{\rm {\rm d}}s}|j(s)\rangle\qquad\qquad\quad\,\,\\
\qquad\qquad+\sum_{n\ne m,j}\left[\rho_{nj}(s)\langle m(s)|\frac{{\rm {\rm d}}}{{\rm {\rm d}}s}|n(s)\rangle-\rho_{mn}(s)\langle n(s)|\frac{{\rm {\rm d}}}{{\rm {\rm d}}s}|j(s)\rangle\right].
\end{split}
\label{Oded}
\end{equation}
Reexpressing the off-diagonal density matrix elements $\rho_{nj}(s)$ and $\rho_{mn}(s)$ in the second line of Eq.~(\ref{Oded}) using Eq.~(\ref{Ode}), and then inserting Eq.~(\ref{Oded}) back into Eq.~(\ref{Ode}), we find that, to the first order of $v$, the terms in the second line of Eq.~(\ref{Oded}) can be dropped. So up to the first order of $v$, we finally have
\begin{equation}
\rho_{mj}(s)=v\frac{\dot{\rho}_{mj}(s)+\left[\rho_{jj}(s)-\rho_{mm}(s)\right]\langle m(s)|\frac{{\rm d}}{{\rm d}s}|j(s)\rangle}{-i\left[E_{m}(s)-E_{j}(s)\right]-\frac{\gamma}{2}\left[A_{m}(s)-A_{j}(s)\right]^{2}}.
\label{PDLPapp0Eq4}
\end{equation}
Further plugging Eq.~(\ref{PDLPapp0Eq4}) into Eq.~(\ref{PDLPapp0Eq3}), we arrive at
\begin{equation}
\begin{split}\rho_{jj}(s_{1})-\rho_{jj}(s_{0})=\sum_{m\neq j}\int_{s_{0}}^{s_{1}}\left\{ v\frac{\dot{\rho}_{mj}(s)\langle j(s)|\frac{{\rm d}H(s)}{{\rm d}s}|m(s)\rangle}{g_{mj}(s)\left[\gamma_{mj}(s)+ig_{mj}(s)\right]}+{\rm c.c.}\right\}{\rm d}s\qquad\qquad\qquad\\
\quad-\sum_{m\neq j}\int_{s_{0}}^{s_{1}}\left\{ v\frac{\left[\rho_{jj}(s)-\rho_{mm}(s)\right]\left|\langle j(s)|\frac{{\rm d}H(s)}{{\rm d}s}|m(s)\rangle\right|^{2}}{g_{mj}^{2}(s)\left[\gamma_{mj}(s)+ig_{mj}(s)\right]}+{\rm c.c.}\right\}{\rm d}s\,,
\end{split}
\label{PDLPapp0Eq5}
\end{equation}
where
\begin{equation}
\gamma_{mj}(s)\equiv\frac{\gamma}{2}\left[A_{m}(s)-A_{j}(s)\right]^{2},\qquad g_{mj}(s)\equiv E_{m}(s)-E_{j}(s).
\label{PDLPapp0Eq6}
\end{equation}

To proceed further, we perform an integration by parts over $\dot{\rho}_{mj}(s)$, the first term on the RHS of Eq.~(\ref{PDLPapp0Eq5}). That term then becomes
\begin{equation}
\begin{split}\sum_{m\neq j}\left.\left\{ v\frac{\rho_{mj}(s)\langle j(s)|\frac{{\rm d}H(s)}{{\rm d}s}|m(s)\rangle}{g_{mj}(s)\left[\gamma_{mj}(s)+ig_{mj}(s)\right]}+{\rm c.c.}\right\}\right|_{s=s_{0}}^{s=s_{1}}
\qquad\qquad\qquad\qquad\\
-\sum_{m\neq j}\int_{s_{0}}^{s_{1}}\left\{\rho_{mj}(s)\left(\frac{{\rm d}}{{\rm d}s}\frac{v\langle j(s)|\frac{{\rm d}H(s)}{{\rm d}s}|m(s)\rangle}{g_{mj}(s)\left[\gamma_{mj}(s)+ig_{mj}(s)\right]}\right)+{\rm c.c.}\right\}{\rm d}s.
\end{split}
\label{PDLPapp0Eq7}
\end{equation}
Note that for a nonzero dephasing rate $\gamma$, all the off-diagonal elements $\rho_{mj}(s)$ would have decayed to zero at the end of a slow pumping cycle. Thus, we can ignore the term proportional to $\rho_{mj}(s_1)$ in Eq.~(\ref{PDLPapp0Eq7}). Using again Eq.~(\ref{PDLPapp0Eq4}), one sees that the term on the second line in Eq.~(\ref{PDLPapp0Eq7}) is at least of order ${\cal O}(v^2)$. Thus this term can also be dropped if we only consider terms up to the first order of $v$. With these clarifications, the only term left in Eq.~(\ref{PDLPapp0Eq7}) is
\begin{equation}
-\sum_{m\neq j}\left.\left\{ v\frac{\rho_{mj}(s)\langle j(s)|\frac{{\rm d}H(s)}{{\rm d}s}|m(s)\rangle}{g_{mj}(s)\left[\gamma_{mj}(s)+ig_{mj}(s)\right]}+{\rm c.c.}\right\} \right|_{s=s_{0}}.
\label{PDLPapp0Eq8}
\end{equation}

The second term in Eq.~(\ref{PDLPapp0Eq5}) can be simplified by replacing $\rho_{jj}(s)$ and $\rho_{mm}(s)$ with their zeroth order expressions, namely, $\rho_{jj}(s_0)$ and $\rho_{mm}(s_0)$. Together with Eq.~(\ref{PDLPapp0Eq8}), we finally have
\begin{equation}
\begin{split}\rho_{jj}(s_{1})=\rho_{jj}(s_{0})-\sum_{m\neq j}\left.\left\{ v\frac{\rho_{mj}(s)\langle j(s)|\frac{{\rm d}H(s)}{{\rm d}s}|m(s)\rangle}{g_{mj}(s)\left[\gamma_{mj}(s)+ig_{mj}(s)\right]}+{\rm c.c.}\right\} \right|_{s=s_{0}}
\qquad\\
-2\sum_{m\neq j}\left[\rho_{jj}(s_{0})-\rho_{mm}(s_{0})\right]\int_{s_{0}}^{s_{1}}v\frac{\gamma_{mj}(s)\left|\langle j(s)|\frac{{\rm d}H(s)}{{\rm d}s}|m(s)\rangle\right|^{2}}{g_{mj}^{2}(s)\left[\gamma_{mj}^{2}(s)+g_{mj}^{2}(s)\right]}{\rm d}s,
\end{split}
\label{PDLPapp0Eq9}
\end{equation}
The expression we show in the main text~[Eq.~($2$) therein] is obtained under the condition $s(t)=vt$. In that case $v$ is a constant and can be pulled out from the above integral in Eq.~(\ref{PDLPapp0Eq9}).

\section{First-order nonadiabatic corrections in two- and three-level systems}
\label{PDLPapp1}

\begin{figure} 
\begin{center}
\resizebox{0.7\textwidth}{!}{%
  \includegraphics[trim=0.6cm 0.7cm 0.5cm 11cm, clip=true, height=!,width=15cm]{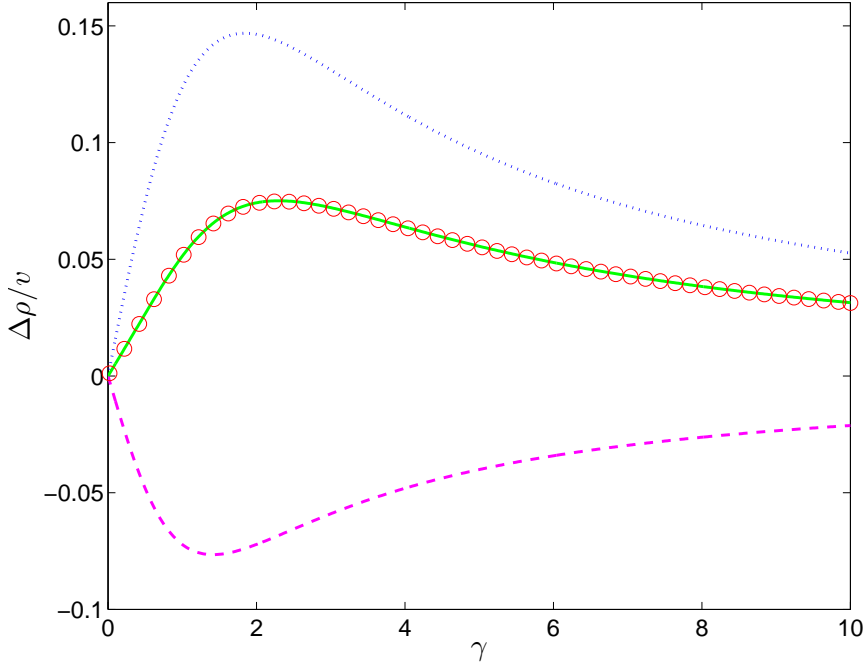}
}
\end{center}
\caption{(color online) Nonadiabatic transition probabilities in LZS under pure dephasing, plotted as a function of the dephasing rate $\gamma$. (magenta)~dashed line: transition probabilities due to initial state coherence~[the first term on the RHS of Eq.~(\ref{FS1})], (blue)~dotted line: transition probabilities due to initial population difference~[the second term on the RHS of Eq.~(\ref{FS1})], (green)~solid line: total transition probabilities in theory, (red)~circles: total transition probabilities obtained by numerically evolving Eq.~(\ref{Leq}). The adiabatic sweeping speed is chosen to be a constant, with $v=10^{-3}$.}
\label{fig:TPC}
\end{figure}

This section is devoted to verifying Eq.~($2$) in the main text or Eq.~(\ref{PDLPapp0Eq9}) here. We first consider a Landau-Zener system~(LZS) subject to Lindblad pure dephasing, as described by Eq.~(\ref{Leq}). The system Hamiltonian is given by $H(s)=\frac{1}{2}(g_0\sigma_x+s\sigma_z)$, where the constant $g_0>0$ equals the minimum energy gap of the instantaneous Hamiltonian $H(s)$.  For convenience we choose $A(s)=H(s)$. As an example, initially~($s=s_0$) the two levels of $H(s_0)$ are assumed to be coherently populated as:
\begin{equation}
\rho(s_0)=\left[\frac{\sqrt{3}}{2}|\psi^{(1)}(s_0)\rangle+\frac{1}{2}|\psi^{(2)}
(s_0)\rangle\right]\left[\frac{\sqrt{3}}{2}\langle\psi^{(1)}(s_0)|+\frac{1}{2}\langle\psi^{(2)}(s_0)|\right],
\label{IS1}
\end{equation}
where $|\psi^{(1)}(s_0)\rangle$ and $|\psi^{(2)}(s_0)\rangle$ denote the ground and excited states of $H(s)$ at $s=s_0$.
In our explicit calculations, we choose $s_0=-1$, $s_1=1$, $g_0=1$. Plugging this initial state into Eq.~(\ref{PDLPapp0Eq9}), the nonadiabatic transition probability up to the first order of $v$ at the final time $s=s_1$ then reads:
\begin{equation}
\Delta\rho=\rho_{22}(s=1)-\rho_{22}(s=-1)=-v\frac{\sqrt{3}\gamma}{8(\gamma^2+2)}
+v\frac{\gamma}{2}\int_{-\pi/4}^{\pi/4}\frac{\cos^4(\theta)}{\gamma^2+4\cos^2(\theta)}{\rm{d}}\theta.
\label{FS1}
\end{equation}
On the RHS of Eq.~(\ref{FS1}), the first term is due to initial state coherence and the second term is an integral~(which can be worked out analytically) over the whole adiabatic process weighted by the population difference of the two levels at $s_0=-1$. To numerically verify Eq.~(\ref{FS1}), we directly evolve the initial state given in Eq.~(\ref{IS1}) using the Lindblad master equation~[Eq.~(\ref{Leq})] from $s_0=-1$ to $s_1=1$. The comparison between theory and numerical results is shown in Fig.~(\ref{fig:TPC}), where an excellent agreement is obtained for a wide range of the dephasing rates.

\begin{figure} 
\begin{center}
\resizebox{0.7\textwidth}{!}{%
  \includegraphics[trim=0.6cm 0.7cm 0.5cm 11cm, clip=true, height=!,width=15cm]{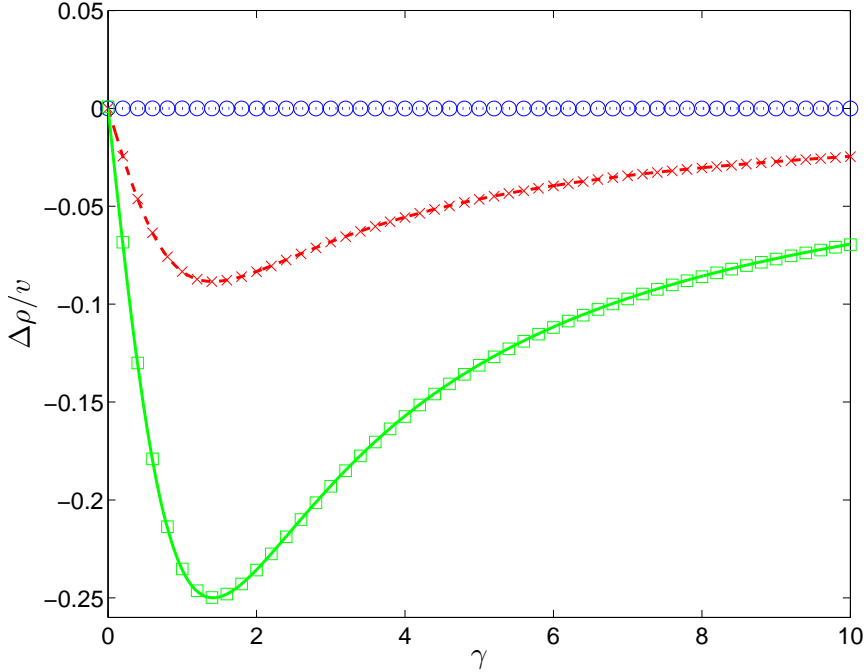}
}
\end{center}
\caption{(color online) Nonadiabatic transition probabilities with respect to the dephasing rate $\gamma$ for three different adiabatic protocols. From top to bottom, the adiabatoc protocols are given by $s=\cos(ut)$, $s=ut$, and $s=1-u^2t^2$, with $u=10^{-3}$, the starting and ending values of $s$ are given by $s_0=-1$ and $s_1=1$. Dotted, dashed, solid lines denote theoretical results, while symbols denote numerical results.}
\label{fig:TPCII}
\end{figure}

To further check how initial state coherence induces corrections to nonadiabatic population transfer, we consider a second initial state:
\begin{equation}
\rho(s_0)=\left[\frac{\sqrt{2}}{2}|\psi^{(1)}(s_0)\rangle+\frac{\sqrt{2}}{2}|\psi^{(2)}
(s_0)\rangle\right]\left[\frac{\sqrt{2}}{2}\langle\psi^{(1)}(s_0)|+\frac{\sqrt{2}}{2}\langle\psi^{(2)}(s_0)|\right].
\label{IS2}
\end{equation}
As seen in Eq.~(\ref{PDLPapp0Eq9}) and also stressed in the main text, the switch-on behavior of an adiabatic process will be important in correcting nonadiabatic transition probabilities in the presence of initial state coherence~(hence important to adiabatic pumping). To check this we investigate three different adiabatic protocols: (I).~$s=ut$, (II).~$s=\cos(ut)$, and (III).~$s=1-u^2t^2$. Assuming that the start times are $t_0=-\pi/u$ for case II and $t_0=-\sqrt{2}/u$ for case III, the sweeping rate of $s$ at the start would be $v=u$, $v=0$ and $v=2\sqrt{2}u$, thus yielding different nonadiabatic corrections. In addition, for this initial state, the two levels of $H(s_0)$ are equally populated in the beginning, so only the first term on the RHS of Eq.~(\ref{PDLPapp0Eq9}) will give rise to nonadiabatic transitions up to the first order in $v$. The theoretical nonadiabatic transition probabilities for these three protocols are respectively given by $\Delta\rho=-u\frac{\gamma}{4(\gamma^2+2)}$, $\Delta\rho=0$, and $\Delta\rho=-2\sqrt{2}u\frac{\gamma}{4(\gamma^2+2)}$. In Fig.~\ref{fig:TPCII} we compare our theory  with numerical results, with excellent agreement for all the three adiabatic protocols.

We have also considered a three-level system described by the Hamiltonian $H(s)=g_0 S_x + s S_z$, where $S_x$ and $S_z$ are spin-$1$ operators. In this case, for a constant sweeping rate $v$ of $s$, the nonadiabatic transition probability on the second excited state is given by
\begin{equation}
\Delta\rho_{11}=-2v\left[\rho_{11}(s_{0})-\rho_{22}(s_{0})\right]D_{1}-vD_{2},
\end{equation}
with
\begin{equation}
D_1=\frac{1}{4} \gamma g_0^2 \left[\frac{\left(2-\gamma^2 g_0 ^2\right) \cot^{-1}(g_0)}{g_0^3}+\frac{\gamma^3\tan^{-1} \left(\frac{\gamma}{\sqrt{\gamma ^2 g_0 ^2+4}}\right)}{\sqrt{\gamma ^2 g_0 ^2+4}}+\frac{2}{g_0 ^4+g_0 ^2}\right]
\end{equation}
and
\begin{equation}
D_{2}=\frac{4\sqrt{2}g_{0}\left\{ \textrm{Re}\left[\rho_{12}(s_{0})\right]\gamma g_{1}-2\textrm{Im}\left[\rho_{12}(s_{0})\right]\right\} \left(g_{1}+2s_{0}-1\right)}{g_{1}^{3}+\left(4+\gamma^{2}g_{1}^{2}\right)
\sqrt{g_{1}^{2}+\left(2g_{1}+1\right)\left(2s_{0}-1\right)}},
\end{equation}
where $\rho_{12}(s_0)$ is the initial off-diagonal density matrix element describing the coherence between the two excited states and $g_{1}\equiv\sqrt{g_{0}^{2}+\left(1-2s_{0}\right)^{2}}$.

\begin{figure} 
\begin{center}
\resizebox{0.9\textwidth}{!}{%
  \includegraphics[trim=0.6cm 1.0cm 0.5cm 2cm, clip=true, height=!,width=15cm]{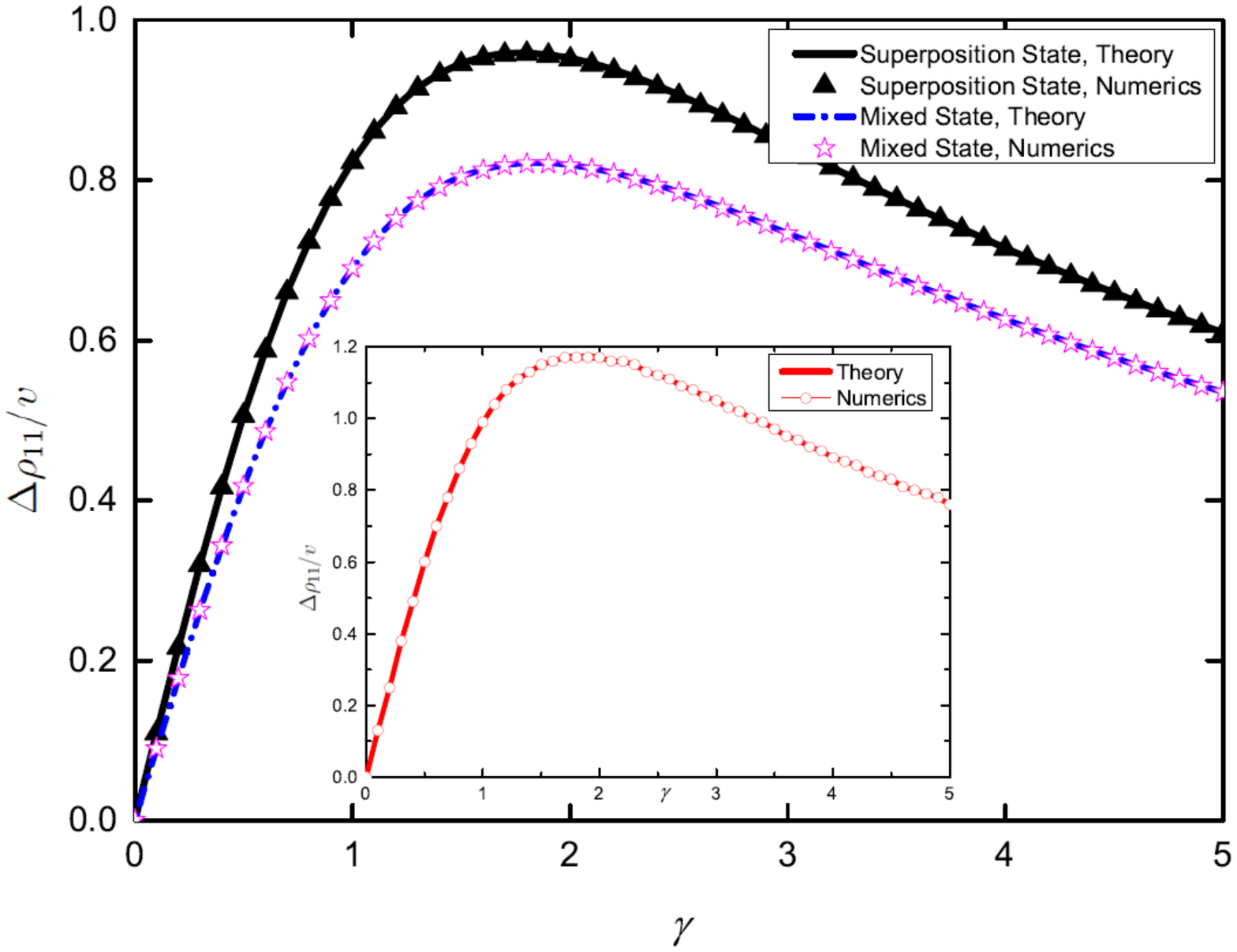}
}
\end{center}
\caption{(color online) Nonadiabatic transition probabilities $\Delta\rho_{11}$ vs. the dephasing rate $\gamma$. For two types of initial states, namely, a superposition state of three instantaneous eigenstates of $H(s_0)$~(hence with initial state coherence) and a mixed state without initial state coherence. In the former case, the initial wavefunction of the system is assumed to be $|\psi(s_0)\rangle=\sqrt{0.8}|1(s_0)\rangle+\sqrt{0.1}|2(s_0)\rangle+\sqrt{0.1}|3(s_0)\rangle$.
In the latter case, the initial density matrix is given by $\rho(s_0)=0.8|1(s_0)\rangle\langle 1(s_0)|+0.1|2(s_0)\rangle\langle 2(s_0)|+0.1|3(s_0)\rangle \langle 3(s_0)|$. In the calculations we set $g_0=1$, $v=10^{-3}$, $s_0=-1$ and $s_1=1$. Inset shows the nonadiabatic transition probabilities if the initial state is prepared in a pure state $|\psi\rangle=|1\rangle$, namely, the second excited state. In all the cases our theory agrees with numerics.}
\label{fig:threelevelplot}
\end{figure}

Figure~\ref{fig:threelevelplot} shows again the agreement between theory and numerics for this case, for a wide range of the dephasing rate $\gamma$ and for various initial states. Comparing the two cases shown in the main panel of Fig.~\ref{fig:threelevelplot}, we indeed see the impact of initial state coherence on nonadiabatic transition probabilities, albeit dephasing.

\section{Detailed Derivations of $f_{ks}$}
\label{PDLPapp2}

As defined in the main text, the number of pumped particles over one adiabatic cycle is given by $Q=\frac{1}{2\pi}\int_{-\pi}^{\pi}{\rm{d}}k\int_{0}^{2\pi}{\rm {d}}s\,f_{ks}$, with $f_{ks}$ defined by
\begin{equation}
f_{ks}\equiv{\rm{Tr}}\left[v^{-1}\rho(s)v_k(s)\right].
\label{PCurrent}
\end{equation}
Here $v_k(s)\equiv \frac{{\rm d} H_k(s)}{{\rm d}k}$ is the group velocity operator. Without loss of generality we assume the adiabatic pumping protocol starts from $s_0=0$ and ends with $s_1=2\pi$. Using
\begin{equation}
f_{ks}=v^{-1}\sum_j\rho_{jj}(s)\langle j(s)|v_{k}(s)|j(s)\rangle + v^{-1}{\sum_{\substack{j,m\\m\ne j}}}\rho_{mj}(s)\langle j(s)|v_k(s)|m(s)\rangle
\end{equation}
as well as the expressions for $\rho_{jj}(s)$ and $\rho_{mj}(s)$ in Eqs.~(\ref{PDLPapp0Eq4}) and~(\ref{PDLPapp0Eq9}), one obtains
\begin{eqnarray}
f_{ks}&=&v^{-1}\sum_{j}\rho_{jj}(0)\frac{\partial E_j(s)}{\partial k}-2{\sum_{\substack{j,m\\m\ne j}}}\left[\rho_{jj}(0)-\rho_{mm}(0)\right]\frac{\partial E_j(s)}{\partial k}\int_{0}^{s}\frac{\big|\langle j(s')|\frac{{\rm{d}}H(s')}{{\rm{d}}s'}|m(s')\rangle\big|^{2}\gamma_{mj}(s')}
{g^2_{mj}(s')\left[\gamma^2_{mj}(s')+g^2_{mj}(s')\right]}{\rm{d}}s'\nonumber\\
&-& 2{\sum_{\substack{j,m\\m\ne j}}}\frac{\partial E_j(s)}{\partial k}{\rm Re}\left\{\rho_{mj}(s')\frac{\langle j(s')|\frac{{\rm d}H(s')}{{\rm{d}}s'}|m(s')\rangle}
{g_{mj}(s')\left[\gamma_{mj}(s')+ig_{mj}(s')\right]}\right\}\bigg|_{s'=0}\nonumber \\
&-& {\sum_{\substack{j,m\\m\ne j}}}\langle j(s)|\frac{{\rm{d}}H(s)}{{\rm{d}}k}|m(s)\rangle\frac{\frac{{\rm d}}{{\rm d}s}\rho_{mj}(s)+\sum_{n}\left[\langle m(s)|\frac{{\rm d}}{{\rm d}s}|n(s)\rangle \rho_{nj}(s)-\rho_{mn}(s)\langle n(s)|\frac{{\rm d}}{{\rm d}s}|j(s)\rangle\right]}{\gamma_{mj}(s)+ig_{mj}(s)}.
\label{ICTsp}
\end{eqnarray}
As emphasized in the main text, all the quantities in the above equation should be understood as functions of quasimomentum $k$, though this dependence is not explicitly spelled out. Let us first focus on the first term on the RHS of Eq.~(\ref{ICTsp}). In principle this term~(inversely proprotional to $v$) can have a large contribution to the pumped number of particles. To highlight other contributions due to nonadiabatic transitions, we assume that both the initial state and the spectrum of $H_k(s)$ and $A_k(s)$ are even functions of $k$. Under this assumption, the first term upon integration over $k$ will have no contribution to the pumping. For the same reason, the second term on the RHS of Eq.~(\ref{ICTsp}) will also vanish. Under this simplification, $f_{ks}$ reduces to
\begin{eqnarray}
f_{ks}=f_{ks}^{(c)}+f_{ks}^{(abd)},
\end{eqnarray}
with
\begin{equation}
f_{ks}^{(c)}=-2{\sum_{\substack{j,m\\m\ne j}}}\frac{\partial E_j(s)}{\partial k}{\rm Re}\left\{\rho_{mj}(s')\frac{\langle j(s')|\frac{{\rm d}H(s')}{{\rm{d}}s'}|m(s')\rangle}{g_{mj}(s')[\gamma_{mj}(s')+ig_{mj}(s')]}\right\}\bigg|_{s'=0}
\end{equation}
and
\begin{equation}
f_{ks}^{(abd)}=-{\sum_{\substack{j,m\\m\ne j}}}\langle j(s)|\frac{{\rm{d}}H(s)}{{\rm{d}}k}|m(s)\rangle\frac{\frac{{\rm d}}{{\rm d}s}\rho_{mj}(s)+\sum_{n}\left[\langle m(s)|\frac{{\rm d}}{{\rm d}s}|n(s)\rangle \rho_{nj}(s)-\rho_{mn}(s)\langle n(s)|\frac{{\rm d}}{{\rm d}s}|j(s)\rangle\right]}{\gamma_{mj}(s)+ig_{mj}(s)}.
\label{ICTsp0}
\end{equation}

Next we focus on the first term on the RHS of Eq.~(\ref{ICTsp0}), which contains $\frac{{\rm d}}{{\rm d}s}\rho_{mj}(s)$.
Because $f_{ks}$ will be under an integration upon $s$ to give the number of pumped particles $Q$, we can consider an integration over $s$ by parts here. Then this term~(under the integration over $s$) will result in two terms,
\begin{equation}
\begin{split}-{\sum_{\substack{j,m\\m\ne j}}}\int_{0}^{2\pi}\frac{\langle j(s)|\frac{{\rm d}H(s)}{{\rm d}k}|m(s)\rangle}{\gamma_{mj}(s)+ig_{mj}(s)}\dot{\rho}_{mj}(s){\rm d}s=-{\sum_{\substack{j,m\\m\ne j}}}\left.\left[\rho_{mj}(s)\frac{\langle j(s)|\frac{{\rm d}H(s)}{{\rm d}k}|m(s)\rangle}{\gamma_{mj}(s)+ig_{mj}(s)}\right]\right|_{s=0}^{s=2\pi}\qquad\\
+{\sum_{\substack{j,m\\m\ne j}}}\int_{0}^{2\pi}\rho_{mj}(s)\left[\frac{{\rm d}}{{\rm d}s}\frac{\langle j(s)|\frac{{\rm d}H(s)}{{\rm d}k}|m(s)\rangle}{\gamma_{mj}(s)+ig_{mj}(s)}\right]{\rm d}s.\,\,\,
\label{PDLPappBeq2}
\end{split}
\end{equation}
Since a nonzero dephasing rate $\gamma$ leads to an exponential decay of all off-diagonal density matrix elements $\rho_{mj}(s)$, the term proportional to $\rho_{mj}(s)$ in Eq.~(\ref{PDLPappBeq2}) is negligibly small at the end of a slow pumping cycle~($s=2\pi$). Thus the only contribution of the first term on the RHS of Eq.~(\ref{PDLPappBeq2}) to the pumping current is due to:
\begin{eqnarray}
{\sum_{\substack{j,m\\m\ne j}}}\left[\rho_{mj}(s)\frac{\langle j(s)|\frac{{\rm d}H(s)}{{\rm{d}}k}|m(s)\rangle}
{\gamma_{mj}(s)+ig_{mj}(s)}\right]\bigg|_{s=0}.
\label{ICTsp1}
\end{eqnarray}
The second term in Eq.~(\ref{PDLPappBeq2}) still involve an integral over $s$. The integration of the off-diagonal density matrix element $\rho_{mj}(s)$ (oscillating with $s$ while decaying) over $s$ indicates that it is at least of the order of $v$, hence negligible as compared to the first term in Eq.~(\ref{PDLPappBeq2}). Alternatively, one may plug our earlier expression for $\rho_{mj}(s)$ in Eq.~(\ref{PDLPapp0Eq4}), only to find that the second term in Eq.~(\ref{PDLPappBeq2}) becomes
\begin{eqnarray}
&-v{\sum_{\substack{j,m\\m\ne j}}}\int_{0}^{2\pi}\frac{\frac{{\rm d}}{{\rm d}s}\rho_{mj}(s)+\sum_{n}\left[\langle m(s)|\frac{{\rm d}}{{\rm d}s}|n(s)\rangle\rho_{nj}(s)-\rho_{mn}(s)\langle n(s)|\frac{{\rm d}}{{\rm d}s}|j(s)\rangle\right]}{\gamma_{mj}(s)+ig_{mj}(s)}\left[\frac{{\rm d}}{{\rm d}s}\frac{\langle j(s)|\frac{{\rm d}H(s)}{{\rm d}k}|m(s)\rangle}{\gamma_{mj}(s)+ig_{mj}(s)}\right]{\rm d}s,
\label{PDLPappBeq3}
\end{eqnarray}
which is indeed of the order of $v$.  As such, $f_{ks}^{(abd)}$ depicted in Eq.~(\ref{ICTsp0}) equivalently~(upon integration of $s$ over one pumping cycle) contains the following contributions that have $v$-independent terms,
\begin{equation}
f_{ks}^{(abd)}=f_{ks}^{(d)}+ f_{ks}^{(ab)},
\end{equation}
where
\begin{equation}
f_{ks}^{(d)}=\frac{1}{2\pi}{\sum_{\substack{j,m\\m\ne j}}}\left[\rho_{mj}(s)\frac{\langle j(s)|\frac{{\rm d}H(s)}{{\rm{d}}k}|m(s)\rangle}{\gamma_{mj}(s)+ig_{mj}(s)}\right]\bigg|_{s=0},
\label{ICTsp1-n}
\end{equation}
and
\begin{equation}
f_{ks}^{(ab)}=-{\sum_{\substack{j,m\\m\ne j}}}\langle j(s)|\frac{{\rm{d}}H(s)}{{\rm{d}}k}|m(s)\rangle\frac{\sum_{n}\left[\langle m(s)|\frac{{\rm d}}{{\rm d}s}|n(s)\rangle \rho_{nj}(s)-\rho_{mn}(s)\langle n(s)|\frac{{\rm d}}{{\rm d}s}|j(s)\rangle\right]}{\gamma_{mj}(s)+ig_{mj}(s)}.
\label{PDLPappBeq5}
\end{equation}
As we have explained earlier, the contributions of those terms involving the off-diagonal density matrix elements to the pumping are at least of order ${\cal O}(v)$ after integrating over $s$. Ignoring these higher-order effects, Eq.~(\ref{PDLPappBeq5}) reduces to
\begin{equation}
f_{ks}^{(ab)}=-{\sum_{\substack{j,m\\m\ne j}}}\langle j(s)|\frac{{\rm d}H(s)}{{\rm d}k}|m(s)\rangle\frac{[\rho_{jj}(s)-\rho_{mm}(s)]\langle m(s)|\frac{{\rm d}}{{\rm d}s}|j(s)\rangle}{\gamma_{mj}(s)+ig_{mj}(s)}.
\label{PDLPappBeq6}
\end{equation}
With the help of $\frac{\langle j(s)|\frac{{\rm d}H(s)}{{\rm d}k}|m(s)\rangle}{g_{mj}(s)}=\langle j(s)|\frac{\rm d}{{\rm d}k}|m(s)\rangle$, this above expression for $f_{ks}^{(ab)}$ becomes
\begin{equation}
f_{ks}^{(ab)}=-{\sum_{\substack{j,m\\m\ne j}}}\frac{\left[\rho_{jj}(s)-\rho_{mm}(s)\right]\langle m(s)|\frac{{\rm d}}{{\rm d}s}|j(s)\rangle\langle j(s)|\frac{{\rm d}}{{\rm d}k}|m(s)\rangle}{\Gamma_{mj}(s)+i},
\label{PDLPappBeq7}
\end{equation}
where $\Gamma_{mj}(s)=\frac{\gamma_{mj}(s)}{g_{mj}(s)}$ is defined in the main text. To proceed further, we shift the derivative of $k$ to act on state $|j(s)\rangle$ and exchange $m$ and $j$ in the summation. With these manipulations, Eq.~(\ref{PDLPappBeq7}) is converted to
\begin{equation}
f_{ks}^{(ab)}={\sum_{\substack{j,m\\m\ne j}}}\rho_{jj}(s)\left[\frac{\langle\frac{{\rm d}}{{\rm d}k}j(s)|m(s)\rangle\langle m(s)|\frac{{\rm d}}{{\rm d}s}j(s)\rangle}{\Gamma_{mj}(s)+i}+\frac{\langle\frac{{\rm d}}{{\rm d}s}j(s)|m(s)\rangle\langle m(s)|\frac{{\rm d}}{{\rm d}k}j(s)\rangle}{\Gamma_{mj}(s)-i}\right].
\label{PDLPappBeq8}
\end{equation}
Recombining the two terms in Eq.~(\ref{PDLPappBeq8}) and using the zeroth order expression for $\rho_{jj}(s)$~(because this only introduces an error proportional to $v$ to the pumping, which can be neglected in a slow pumping protocol), we arrive at
\begin{equation}
f_{ks}^{(ab)}=f_{ks}^{(a)}+f_{ks}^{(b)},
\end{equation}
where
\begin{equation}
f_{ks}^{(a)}=2{\sum_{\substack{j,m\\m\ne j}}}\rho_{jj}(0)
\frac{\Gamma_{mj}(s){\rm Re}\left[\langle \frac{{\rm d}}{{\rm d}k} j(s)|m(s)\rangle\langle m(s)|\frac{{\rm d}}{{\rm d}s}j(s)\rangle\right]}{\Gamma^2_{mj}(s)+1}
\end{equation}
and
\begin{equation}
f_{ks}^{(b)}=2{\sum_{\substack{j,m\\m\ne j}}}\rho_{jj}(0)\frac{{\rm Im}\left[\langle \frac{{\rm d}}{{\rm d}k} j(s)|m(s)\rangle\langle m(s)|\frac{{\rm d}}{{\rm d}s}j(s)\rangle\right]}{\Gamma^2_{mj}(s)+1}.
\end{equation}

Combining all these results together, we finally have $f_{ks}=f_{ks}^{(a)}+f_{ks}^{(b)}+f_{ks}^{(c)}+f_{ks}^{(d)}$, which is precisely the result used in the main text. It should be emphasized that all the four $v$-independent subterms of $f_{ks}$ are derived based on nonadiabatic corrections captured to the first order of $v$.


\begin{thebibliography}{99}
\bibitem{notesupp} For numerical checking of our derived results regarding the population dynamics during adiabatic following with dephasing, as well as detailed derivations of the four subterms of $f_{ks}$, see Appendix.
\end{thebibliography}
\end{document}